# Antisite disorder driven spontaneous exchange bias effect in La$_{2-x}$Sr$_x$CoMnO$_6$ (0 ≤ x ≤ 1)

J. Krishna Murthy[1], K. D. Chandrasekhar[2], H. C. Wu[2], H. D. Yang[2], J. Y. Lin[3] and A. Venimadhav[1]

[1]*Cryogenic Engineering Centre, Indian Institute of Technology, Kharagpur-721302, India.*

[2]*Department of Physics and Center for Nanoscience and Nanotechnology, National Sun Yat-Sen University, Kaohsiung 804, Taiwan.*

[3]*Institute of Physics, National Chiao Tung University, Hsinchu 30010, Taiwan.*

## Abstract

Doping at the rare-earth site by divalent alkaline-earth ions in perovskite lattice has witnessed a variety of magnetic and electronic orders with spatially correlated charge, spin and orbital degrees of freedom. Here, we report an antisite disorder driven spontaneous exchange bias effect as a result of hole carrier (Sr$^{2+}$) doping in La$_{2-x}$Sr$_x$CoMnO$_6$ (0 ≤ $x$ ≤ 1) double perovskites. X-ray diffraction and Raman spectroscopy have evidenced an increase in disorder with the increase of Sr content up to x = 0.5 and thereby decreases from $x$ = 0.5 to 1. X-ray absorption spectroscopy has revealed that only Co is present in mixed valent Co$^{2+}$ and Co$^{3+}$ states with Sr doping to compensate the charge neutrality. Magnetotransport is strongly correlated with the increase of antisite disorder. The antisite disorder at the B-site interrupts the long-range ferromagnetic order by introducing various magnetic interactions and instigates reentrant glassy dynamics, phase separation and canted type antiferromagnetic behavior with the decrease of temperature. This leads to novel magnetic microstructure with unidirectional anisotropy that causes spontaneous exchange bias effect that can be tuned with the amount of antisite disorder.

# I. Introduction

A strong interplay among the charge, lattice, orbital and spin degrees of freedom in perovskite materials ($ABO_3$, A-rare-earth and B-transition metal ions) induces distinct fascinating, complex and richness of the physical phenomena like, metal-insulator transition, colossal magnetoresistance, superconductivity, charge/orbital ordering, and multiferroicity [1-7]. One of the straightforward experimental methods of tuning these extraordinary physical properties is by doping at the A-site (with cations of different charge/radii). The quenched disorder with local distortion arising from the difference in ionic radii at the A-site cation and/or random Columbic potentials with the multiple valence states [8] is a provoking agent for the suppression of ordering parameters like, magnetic, charge ordering and superconductivity [9-11], however, it induces interesting properties like, multiglass behavior, phase separation ferroelectricity, and exchange bias (EB) effect [12-15]. The energy balance between the competing phases leads to the phase coexistence at submicron length scales and induce meta-magnetic/electric phase transitions [16, 17].

Double perovskite $La_2(Co/Ni)MnO_6$ systems have attracted considerable interest in recent years due to their magnetoelectric (ME) effect and possible applications in spintronics. Structural and magnetic studies have demonstrated the long-range FM ordering that originates from the superexchange interaction between $(Co/Ni)^{2+}$ and $Mn^{4+}$ magnetic species arranged with the rock-salt configuration at the B-sublattice of the perovskite cell. However, the existence of antisite disorder (ASD) is inevitable in double perovskite structure and plays a vital role on their physical properties. In FM/ferrimagnetic double perovskites, this ASD induced AFM exchange interactions are responsible for the magnetic frustration, phase separation and large low field magnetoresistance, further; it also reduces the saturation magnetization and destroy the half-metallicity [18, 19]. On the other hand, a large ME coupling over a broad temperature range was found in partially disordered $La_2NiMnO_6$ and such disorder manifest a re-entrant spin glass (RSG) behaviour at low temperatures, while fully ordered sample has shown feeble ME effect [20, 21]. Contrastingly, a large ME was observed in the highly ordered isostructural $La_2CoMnO_6$ (LCMO) sample in single crystals and polycrystalline forms, while relatively small ME response was reported in the disordered sample [22-24]. Interestingly, a divalent cation ($Sr^{2+}$) doping at the rare-earth site in $La_2NiMnO_6$ has led to the magnetic disorder and exhibited EB effect [25].

In magnetically phase separated systems, the EB effect is anticipated. EB is an interface magnetic coupling phenomena which is manifested as the hysteresis loop shift along the field axis after cooling the system under magnetic field. EB effect is ubiquitous to spintronic applications, hence, understanding and controlling of this effect with the disorder is essential. Lately, there has been a great interest in electrical field control of EB devices [26]. In certain systems below the blocking temperature a spontaneous loop shift can be observed without the assistance of external magnetic field in cooling mode and this unusual zero-field-cooled (ZFC) M (H) loop shift is called zero-field-cooled EB (ZEB) or spontaneous EB effect [27-29]. Such a spontaneous EB effect will be of great interest in the case of electric field control of EB devices as it eliminates the requirement of external magnetic field to create the unidirectional anisotropy. Recently, we have reported a giant value of spontaneous and conventional EB effects in $La_{1.5}Sr_{0.5}CoMnO_6$ system [30]. Metamagnetic behavior and a field induced phase separation below canted antiferromagnetic (CAF) transition are found to be responsible for the observed giant ZEB and conventional exchange bias (CEB) effects. To unveil the reason behind the complex magnetic behavior and field induced unidirectional anisotropy in the phase separated region, we have investigated the effect of Sr doping on the EB phenomena. Our study signifies the spin disorder to order state with Sr doping and as a consequence of ASD, a novel magnetic interface is formed that sets ZEB effect.

The paper is structured as following sections: section-II describes the details of various experiments employed to characterize the samples. Section-III describes the preparation details of $La_{2-x}Sr_xCoMnO_6$ ($0 \leq x \leq 1$) samples, structural characterization by the x-ray diffraction (XRD), x-ray absorption spectra (XAS) and Raman studies. The effect of Sr induced disorder on magnetotransport behavior is detailed in section-IV. Further, the temperature and magnetic field dependent dc and ac susceptibility studies are investigated and presented in a phase diagram in section-V. Finally in section-V, importance of ASD is summarized.

## II. Experimental details:

Polycrystalline $La_{2-x}Sr_xCoMnO_6$ ($0 \leq x \leq 1$) bulk samples were prepared by conventional sol-gel method and their synthesis details were given in Ref. [22]. Obtained precursor powder was calcinated at 1300 $^o$C for 24 h. Crystal structural analysis was done using high resolution x-ray diffraction (HRXRD) with Cu-K$\alpha$ radiation. For the electronic structural study, we have carried out XAS of Co-$L_{2,3}$ and Mn-$L_{2,3}$ and the data was collected at the Dragon beam line of the National Synchrotron Radiation Research Centre in Taiwan with energy resolution of 0.25 eV at the Co-$L_3$ edge (~780 eV). Raman spectra data was recorded in the 180$^o$ back scattering geometry using a 514 nm excitation of air-cooled Argon Ion laser (Renishaw InVia Reflex Micro-Raman Spectrometer). Laser power at the sample was ~10 mW and typical spectral acquisition time was ~2 minutes with the spectral resolution of 1 cm$^{-1}$. Temperature and magnetic field dependent dc and ac susceptibility magnetic measurements were carried out by using Quantum Design SQUID-VSM magnetometer. Temperature dependent electrical resistivity under magnetic field was measured with the conventional four-probe method in closed cycle cryogen-free superconducting magnet system.

## III. Synthesis and structural studies of $La_{2-x}Sr_xCoMnO_6$ ($0 \leq x \leq 1$) samples

**(A) Synthesis and crystal structural study**

The XRD pattern of La$_{2-x}$Sr$_x$CoMnO$_6$ (LSCMO) ($0 \leq x \leq 1$) series of samples is displayed in the **Fig. 1(a)**. In comparison with the undoped sample, the Sr doped samples for $x = 0.1$ to 0.5 have shown peak splitting at $2\theta \sim 32.5°$. With further increase in $x$, the XRD pattern showed a single peak as shown in the **Fig. 1(b)**. For $x > 1$ we have observed additional peaks corresponds to Co$_3$O$_4$ and MnO$_2$ phases (not shown here). The parent LCMO bulk sample is refined with monoclinic crystal structure in P2$_1$/n space group while $x = 0.1$ sample has been refined to mixed crystallographic phases having monoclinic and disordered rhombohedra ($R\overline{3}c$) structures. Sr doping for $x = 0.25$ to 0.5 range have showed only disordered rhombohedral phase and the over doped ($x = 0.6$ to 1.0) samples refined well with mixture of disordered rhombohedral ($R3c$) and cubic ($Fm\overline{3}m$) phases. The XRD pattern with Rietveld refinement for selected samples ($x = 0$ and 0.5) is shown in **Fig. 1(c) & (d).** The obtained crystal structure, lattice parameters and Wyckoff positions of all the samples are listed in **Table 1**. The average bond lengths of the transition metal cations with neighboring oxygen atoms in the (Co/Mn)O$_6$ octahedron were obtained from the Diamond software. It is found that Mn-O bond length of $R\overline{3}c$ phase does not change after Sr doping while the bond length of Co-O shrinks continuously up to $x = 0.5$. This indirectly suggests the mixed valence states of Co ions with Sr doping while the valence state of Mn$^{4+}$ remains independent of doping. As shown in **Table 1**, the lattice parameter '**a**' decreases continuously with Sr doping. This lattice compression is consistent with the smaller ionic radius of Co$^{3+}$ (ionic radii of Co$^{3+} \sim 75$ pm and for Co$^{2+}$ it is $\sim 88.5$ pm) [28,29].

**(B) X-ray absorption spectra:**

To confirm the valence states of Co and Mn, we have measured elemental and site specific x-ray absorption near edge spectra (XANES) and **Fig. 2 (a) & (b)** shows the room temperature normalized XANES measured in total electron yield (TEY) mode for the series of LSCMO samples. The valence state of Mn is compared with the reference spectra of MnO (Mn$^{2+}$), Mn$_2$O$_3$ ( Mn$^{3+}$) and MnO$_2$ (Mn$^{4+}$) [31-33] as shown in the **Fig. 2(a)**. These spectral features of Mn -$L_{2,3}$ edges of all the samples matches well with that of MnO$_2$ spectra, and this confirms +4 valence state of Mn and it is consistent with the unchnaged bond length of Mn-O as obtained from XRD. For Co case, in undoped LCMO sample, as shown in **Fig. 2 (b)**, the observed peak position of $L_3$ edge at 779.6 eV and spectral shape matches well with that of CoO standard spectra of Ref. [31] and this confirms +2 valence state of Co. With Sr doping the Co$^{2+}$ peak gets suppressed and the trivalent state of Co (i.e., Co$^{3+}$) peak appears at $\sim$780.8 eV and the observed XANES spectra for $x = 1$ sample matches with that of Sr$_2$CoO$_3$Cl standard spectra, where Co resides in 3+ state [31, 34, 35]. From **Fig. 2(b)** it is clear that the valence state of cobalt ions increases with Sr doping from Co$^{2+}$ to Co$^{3+}$. In order to elucidate on the spin state of Co$^{3+}$ (either in high spin with 3d$^6$, S = 2 or low spin, 3d$^6$, S = 0), the spectra is compared with the XANES spectra of Sr$_2$CoO$_3$Cl for the high spin state and EuCoO$_3$ for the low spin state [31, 34]. The XANES of EuCoO$_3$ in low spin state is characterized with the main peak followed by a shoulder at higher energies in both $L_2$ and $L_3$-edge. Contrastingly, in our Sr doped samples the shoulder is present at lower energies than the main peak in $L_3$-edge and this is similar to the high spin state of Sr$_2$CoO$_3$Cl spectra [34].

## (C) Raman spectra study:

**Fig. 3(a) & (b)** shows the room temperature Raman spectra of LCMO and La$_{2-x}$Sr$_x$CoMnO$_6$ ($0 < x \le$ 1) samples respectively. The observed Raman spectra are consistent with the previous reports where two peaks associated with stretching mode (A$_{1g}$) and mixed mode B$_{1g}$ (anti-stretching and vibration) can be noticed [36, 37]. Sr doping changes the position, symmetric nature and intensity of both the peaks. **Fig. 3(c) & (d)** shows the variation of Raman shift of the two broad peaks at A$_{1g}$ ~ 645 cm$^{-1}$ and B$_{1g}$ ~ 490 cm$^{-1}$ with Sr doping and these values are obtained from the Lorentzian fit to the spectra. Here, the A$_{1g}$ peak shifts to lower wavenumber side (softening) for $x = 0$ to 0.5, while it shifts towards higher wavenumber side (hardening) for the $x = 0.6$ to 1.0 samples. On the other hand the B$_{1g}$ mode shows exactly opposite trend with Sr doping. Another important parameter obtained from the Raman spectra is full width at half maxima (FWHM), a measure of the phonon lifetime that in turn depends on the various factors such as: (i) disorder present in the system and (ii) biphasic crystal structure in the sample [36, 38]. We have plotted the variation of FWHM corresponding to both A$_{1g}$ and B$_{1g}$ modes with Sr content as shown in **Fig. 3(c) & (d)** respectively. Here, with the increasing of Sr doping FWHM is found to be maximum for $x = 0.5$. This is consistent with the structural data.

## (IV) Temperature and magnetic field dependent transport behavior

**Fig. 4(a)** shows the temperature-dependent electrical resistivity ($\rho$) of Sr doped La$_{2-x}$Sr$_x$CoMnO$_6$ ($x = 0, 0.1, 0.25, 0.4, 0.5, 0.6, 0.75$, and 1) samples. Here, all the compositions have exhibited the semiconducting behavior, i.e., d$\rho$/dT< 0 and the resistivity of the samples decrease monotonically with the carrier doping [39-42]. However, there is no anomaly in resistivity near to magnetic ordering at T$_C$ that suggests the electron mobility of the system is controlled by thermal energy rather than magnetic ordering. The enhanced electrical conduction with doping can be explained based on the new conduction path of Co$^{3+}$-O$^{2-}$-Mn$^{4+}$ within the matrix of the superexchange interactions among the various magnetic species [43]. A large magnetoresistance (MR) of ~ 31% was reported in ordered parent LCMO single crystals and bulk samples [23, 44]. Though the origin of large MR is not well understood, an enhanced spin transport due to the alignment of neighboring transition metal ions with a magnetic field has been considered as a possible reason [44]. The isothermal magnetic field dependence of MR with Sr doping at 110 K is shown in the **Fig. 4(b).** Here, a maximum MR of the parent LCMO is 34 %, similar to the single crystals data, and this value increases to 51 % for $x = 0.5$, and then decrease for $x \ge 0.6$. Such a non-monotonic variation of MR with Sr doping (as shown in the inset to **Fig. 4(b)**) suggests the strong correlation of magnetotransport with the disorder.

## V. Magnetization study

### (A) Temperature dependent dc susceptibility:

Temperature dependent magnetization of $La_{2-x}Sr_xCoMnO_6$ samples in zero-field cooled (ZFC) and field-cooled warming (FCW) modes for 100 Oe dc field is shown in the **Fig. 5(a-f)**. Here, the parent compound LCMO shows a single magnetic transition at ~ 230 K and it has been assigned to the $Co^{2+}$-$O^{2-}$-$Mn^{4+}$ FM superexchange interactions [36, 45]. With Sr doping from $x = 0.1$ to 0.4, in addition to high-temperature FM phase ($T_{C1}$) another dominant magnetic ordering at $T_{C2}$ can be observed. This second magnetic transition can be attributed to $Co^{3+}$-$O^{2-}$-$Mn^{4+}$ FM superexchange interactions [30]. Further, the magnetic anomaly around 90-105 K in both FCC and ZFC magnetization has been assigned to a glassy like behavior and will be discussed later. At temperatures around 40-50 K depending on the Sr doping level ($x = 0.1$ to 0.75) a phase separation (PS) state containing FM and glassy phases can be noticed and further below, a CAF phase is established [30]. The PS temperature ($T_P$(K)) is estimated from the first derivative of $M_{ZFC}$ with respect to temperature as shown in the inset of **Fig. 5(b-f)**. For doping $x > 0.5$, the magnetic glass anomaly is suppressed while the FM interactions corresponding to $Co^{3+}$-$O^{2-}$-$Mn^{4+}$ gets enhanced. Correspondingly, the PS state is shifted to low temperature (~10 -15 K) in $x = 0.75$ and vanishes for $x = 1$ sample. Further, one can notice that the magnitude of the magnetization decreases with the increase of Sr doping. From the **Fig. 5**, it is found that the magnetic irreversibility temperature ($T_{irr}$) (i.e., bifurcation in between FCC and ZFC magnetization) coincides with the paramagnetic (PM) to FM transition in $x = 0$ to 0.5 samples. Such a bifurcation in $x = 0.1$ to 0.5 can be attributed to the frustration leading to glassy phase. While for $x > 0.5$, $T_{irr}$ is present even at temperatures well above the magnetic ordering ($T_{C2}$), and this suggest the presence of short-range FM interactions [46].

### (B) Temperature and frequency dependent ac susceptibility study:

**Fig. 6(a-f)** shows the temperature dependent in-phase component ($\chi'$) of ac susceptibility on LSCMO samples. In the parent LCMO compound the frequency independent peak in $\chi'$(T) corresponds to the FM transition. In the doped samples, observed multiple peaks at $T_{C1}$ and $T_{C2}$ have also showed frequency independent nature consistent with FM transition. The variation of $T_{C1}$, $T_{C2}$ (obtained from the ac susceptibility data) and $T_p$ and $T_{CAF}$ (obtained from $M_{ZFC}$ data) with Sr doping are listed in **Table 2**. A clear frequency dependence of the peak at ~ 95-105 K ($T_f$) in doped samples and its shift to higher temperature with the increase of frequency suggests the reentrant glassy dynamics. However, ac susceptibility shows no signature of CAF ordering at low temperatures. As shown in the inset of **Fig. 6(b-e)**, in doped samples, a shift of $T_f$ with relaxation time ($\tau$) has been analyzed by the critical slowing down power law [30, 47, 48]. The $T_f$ vs. $\tau$ data fits well to the power law $\tau = \tau_0 (\frac{T_f - T_g}{T_g})^{-zv}$ as shown in the inset to **Fig. 6(b-e)**, where $\tau_0$ is the microscopic spin relaxation time, $T_g$

is glassy freezing temperature and $zv$ denotes the critical exponent. From the fitting, the obtained $T_g$, $\tau_0$ and $zv$ values with Sr are listed in the **Table 3**. The high $\tau_0$ (~$10^{-6}$ - $10^{-7}$ sec) and small value of $zv$ (3 - 6) in the case of $x = 0.1$ to 0.4 indicates the freezing of magnetic clusters rather than the individual atomic spins suggesting the presence of cluster glass (CG) like behavior. In case of $x = 0.5$ sample; $\tau_0$ ~ $4.23 \times 10^{-11}$ sec and $zv = 10.34$, reveal the RSG nature [30, 47]. Further, the double dip memory and aging effects are the characteristic features of spin glass (SG) phase and were confirmed in $x = 0.5$ doped sample in our previous work [30].

Corroborating the dc and ac susceptibility measurements with structural data, a phase diagram for La$_{2-x}$Sr$_x$CoMnO$_6$ ($0 \leq x \leq 1$) series of samples is shown in the **Fig. 7**. It is clear that the magnetic glassy behavior in the temperature regime of ~ 50 -110 K takes a sudden change from CG (x = 0.1 to 0.4) to SG state at $x = 0.5$ where Co$^{2+}$ and Co$^{3+}$ are present in equal amounts. The various competing magnetic exchange interactions like, Co$^{2+}$-O$^{2-}$-Mn$^{4+}$ (FM), Co$^{3+}$-O$^{2-}$-Mn$^{4+}$ (FM), Co$^{3+}$-O$^{2-}$-Co$^{3+}$ (AFM), and Co$^{2+}$-O$^{2-}$-Co$^{2+}$ (AFM) with large magnetic ASD drives the system to SG state [30, 49]. And complete absence of glassy nature is realized for higher doping i.e., $x > 0.5$. And the end members will have all Co ions in either 2+ state (for x = 0) or 3+ state (for $x = 1$) and will have one defined FM ordering. This indicates that Sr doping induces spin disorder to order state in LCMO.

### (D) Isothermal field-dependent magnetization study:

**Fig .8(a-e)** shows the field variation of magnetization, M (H) curves at 5 K for the selected samples ($x = 0, 0.25, 0.5, 0.75$ and 1) in two modes i.e., ZFC and field cooled with 5 T. Here, the parent compound ($x = 0$) exhibits a well-defined hysteresis loop with large remnant magnetization (M$_r$) and shows saturation like behavior for fields $\geq$ 5 T as depicted in **Fig. 8 (a)**. The obtained high magnetization (M$_S$) ~ 5.75 $\mu_B$/f.u. at 6 T is close to the theoretically calculated spin only value of 6 $\mu_B$/f.u. as expected for the FM alignment of Co$^{2+}$-O$^{2-}$-Mn$^{4+}$ magnetic species [45]. From the **Fig. 8(b-e)**, it is clear that Sr doping shows a significant effect on the shape of the hysteresis loop and magnetization value. **Fig. 8(f)** shows the Sr content variation of M$_S$ obtained at 6 T, M$_r$ and coercive field (H$_C$) values at 5 K. Both M$_S$ and M$_r$ decreases with increasing Sr up to $x = 0.5$ and then shows increasing trend for $x > 0.5$, while H$_C$ variation with Sr doping shows an opposite trend and is consistent with the variation in magnetic disorder. With Sr doping the magnetization value decreases and reaches lowest for $x = 0.5$ and beyond this doping magnetization property improves. A large value of H$_C$ and low values of M$_r$ and M$_S$ at 5 K in $x = 0.5$ sample supports the presence of more magnetic disorder. Here, we estimated ASD which is the main cause for the reduction in M$_S$ values by using [50],

$$M_S = (1-2ASD) [M_{Co}+M_{Mn}] +x (2ASD-1),$$

Here '$x$' denotes the amount of Sr doping and $M_{Co}$ & $M_{Mn}$ are the theoretically calculated spin only magnetic moments of Co and Mn ions respectively. In this expression, the first term indicates the contribution of ASD and second term suggests the reduction in $M_S$ due to hole-carrier doping. Accordingly, we have calculated % of ASD with Sr variation and is plotted as shown in the **Fig. 8 (f)**. Maximum ASD is found in $x = 0.5$ and it matches well with the structural, Raman and magnetotransport data. Here, $M_S$ has shown an almost linear dependence on the ASD.

**(E) ZEB and CEB effects in La$_{2-x}$Sr$_x$CoMnO$_6$ samples**

A definite M (H) loop shift at 5 K (as shown in **Fig. 8**(a)-(e)) in both ZFC and FC modes for doped samples indicates asymmetry in hysteresis loop about the origin in ZFC as well as in FC modes and they illustrate the corresponding ZEB and CEB effects. In the parent sample, either of these effects is found to be absent. While Sr doped samples exhibited ZEB effect for $x = 0.1$ to 0.75. This loop shift is enhanced further in cooling the samples under FC mode and much higher EB shifts are obtained in CEB effect. The loop asymmetric along the field axis and magnetization axis can be quantified as EB field ($H_{EB} = (|H_{C1}|-|H_{C2}|)/2$) and EB magnetization ($M_{EB} = |M_{r1}|-|M_{r2}|)/2$) respectively. Here, $H_{C1}$ and $H_{C2}$ are the positive and negative intercepts of the magnetization curve with field axis and $M_{r1}$ and $M_{r2}$ are the positive and negative intercepts of M (H) curve with magnetization axis respectively. Obtained values of ZEB and CEB effects with Sr doping are are shown in **Fig. 9.** Like ZEB effect, CEB also increases with Sr content from $x = 0.1$ to 0.5 and then becomes maximum for $x = 0.5$ and then decreases for higher Sr.

From the M (T) and M (H) data it is clear that the coexistence of FM and CAF phases appear in La$_{2-x}$Sr$_x$CoMnO$_6$ samples for a broad range of Sr ($0 \leq x \leq 1$) doping. CAF and PS are important ingredients to obtain ZEB [30]. At low temperatures < 10 K, the field induced metamagnetic phase transition from CAF to FM phase is responsible for the exchnage bias effect. With ZFC, the system undergoes to CAF ordering at low temperature. During the initial magnetization, for magnetic field strength higher than the critical field which depends on temperature, induces FM phase. While decreasing field (in the second cycle of M (H) loop) the field induced FM phase is kinetically arrested and coexists with CAF matrix that creates a large unidirectional anisotropy at their interface. In FC case, there exist FM clusters even at H = 0 Oe below the CAF transition that leads to giant CEB effect [30]. A close look reveals that below CAF transition, the CEB is actually an enhanced effect of the ZEB. This explanation is valid for all other Sr doped samples where ZEB is present. A small CEB is observed in between CAF and PS regions, the unidirectional anisotropy formed at the FM and SG interface is responsible for the CEB effect and this is in a way similar to EB effect in the phase separated cobaltates and manganites [51].

Generally in phase separated systems, the observed EB effect can be explained qualitatively based on the Meiklejohn-Bean (MB) model [52]. According to this model under certain assumptions [51, 53], the amount of EB field from the hysteresis loop shift in inhomogeneous magnetic systems can be estimated as, $H_{EB} = \frac{2\sqrt{A_{AFM}K_{AFM}}}{t_{FM}M_{FM}}$ here, $A_{AFM}$ and $K_{AFM}$ are the exchange stiffness and uniaxial anisotropy energy of AFM phase respectively, $t_{FM}$ and $M_{FM}$ are the thickness and saturation magnetization of FM layer respectively. In present case, with Sr doping for $x = 0.1$ to $x = 0.5$, the volume fraction of CAF phase increases and is responsible for the decreasing of $M_S$ values (as shown in the **Fig. 8(f)**). Such low temperature CAF anisotropy and the coupling strength at FM/AFM is the possible origin for the increase of EB effect in both ZFC and FC modes (**Fig. 9**) and correspondingly both ZEB and CEB effects increases and becomes maximum for $x = 0.5$ with high ASD. Further, for higher doping of $x > 0.5$, the weakening of AFM correlations and the increasing of average size of FM clusters analogues to $t_{FM}$ and $M_{FM}$ can reduce the interface coupling strength and unidirectional anisotropy; consequently it decreases the resultant ZEB and CEB effects.

**(6) Conclusions**

Crystal structure of $La_{2-x}Sr_xCoMnO_6$ is sensitive to the Sr doping and Raman results suggest that the disorder increases with Sr doping and maximum for $x = 0.5$, further doping leads towards the order state. We can summaries that hole doping increases the ASD and various AFM interactions that systematically destroy long range magnetic ordering and induce magnetic glass and PS state with CAF ordering at low temperatures. Our results demonstrate the observation of such complex magnetic behavior in $La_{2-x}Sr_xCoMnO_6$ ($0 \leq x \leq 1$) samples and observed field induced novel-magnetic interface related ZEB phenomena. The study signifies the impact of ASD on magnetic and transport properties and presented large values of ZEB, CEB and MR. Importantly tuning of EB with ASD disorder can be a constructive approach for designing new materials for spintronic applications.

**Acknowledgments:**

The authors acknowledge DST and IIT Kharagpur, India for funding VSM-SQUID magnetometer and FIST grant for establishing cryogen-free high magnetic field facility. Krishna thanks to CSIR-UGC, Delhi for SRF.

**Table 1:** *The structural parameters of Sr doped La$_{2-x}$Sr$_x$CoMnO$_6$ (0 ≤ x ≤ 1) samples estimated from the Rietveld refinement.*

| La$_{2-x}$Sr$_x$CoMnO$_6$ | x = 0 | x = 0.1 | x = 0.25 | x = 0.4 | x = 0.5 | x = 0.75 | x = 1.0 |
|---|---|---|---|---|---|---|---|
| *Crystal structure* | mono clinic | monoclinic + disordered rhombohedra | disorder ed rhombo hedra | disordered rhombohe dra | disordered rhombohe dra | disordered rhombohedra + cubic | disordered rhombohedra + cubic |
| *Space group* | P 2$_1$/n | P 2$_1$/n + R3c | R 3c | R 3c | R 3c | R 3c+ Fm-3m | R 3c + Fm-3m |
| *a (Å):* | 5.5223 | 5.4812+5.5124 | 5.4969 | 5.4825 | 5.4717 | 5.4785+7.6857 | 5.467+7.6717 |
| *b (Å):* | 5.4862 | 5.4776+5.5124 | 5.4969 | 5.4825 | 5.4717 | 5.4785+7.6857 | 5.467+7.6717 |
| *c (Å):* | 7.7730 | 7.7873+13.2547 | 13.2516 | 13.2515 | 13.2510 | 13.275+7.6857 | 13.284+7.6717 |
| *α, β and γ (degree):* | 90.04 | β =90.72 & α=β =90 γ=120 | α=β =90 γ=120 | α=β =90 γ=120 | α=β =90 γ=120 | α=β =90 γ=120 & α=β= γ=90 | α=β =90 γ=120 & α=β= γ=90 |
| *Bond length Co-O Mn-O (degree)* | 2.0442 1.9402 | 1.9634 1.9361 | 1.9527 1.9509 | 1.9447 1.9512 | 1.9409 1.9508 | 1.9402 1.9511 | 1.9403 1.9509 |
| $\chi^2$ | 1.3 | 1.4 | 1.2 | 1.3 | 1.2 | 1.840 | 2.6 |

**Table 2**: *High temperature FM (T$_{C1}$) and low temperature FM (T$_{C2}$) ordering, magnetic glassy (T$_g$), PS temperature (T$_p$) and canted AFM (T$_{CAF}$) for Sr doped La$_{2-x}$Sr$_x$CoMnO$_6$ (0 ≤ x ≤ 1) samples.*

| La$_{2-x}$Sr$_x$CoMnO$_6$ | T$_{C1}$ (K) | T$_{C2}$ (K) | T$_g$ (K) | T$_p$ (K) | T$_{CAF}$ (K) |
|---|---|---|---|---|---|
| x = 0.0 | 230 | ____ | ____ | ____ | |
| x = 0.1 | 221 | 156 | 101.31 | 40 | 7 |
| x = 0.25 | 215 | 152 | 94.77 | 45 | 8 |
| x = 0.4 | 167 | 148 | 92.31 | 48 | 10 |
| x = 0.5 | ____ | 177 | 90.13 | 50 | 12 |
| x = 0.6 | ____ | 150 | ____ | 25 | 11 |
| x = 0.75 | ____ | 153 | ____ | 15 | 10 |
| x = 1.0 | ____ | 175 | ____ | ____ | ____ |

**Table 3**: *List of fitting parameters: freezing temperature (T$_g$), relaxation time ($\tau$) and critical exponent (zv) for the Sr doped La$_{2-x}$Sr$_x$CoMnO$_6$ (0.1 $\leq$ x $\leq$ 0.5) samples*

| La$_{2-x}$Sr$_x$CoMnO$_6$ | T$_g$ (K) | $\tau_0$ (sec) | zv |
|---|---|---|---|
| x = 0.1 | 101.31 | 6.31x10$^{-7}$ | 3.09 |
| x = 0.25 | 94.77 | 4.54x10$^{-7}$ | 4.94 |
| x = 0.4 | 92.31 | 1.71x10$^{-7}$ | 6.02 |
| x = 0.5 | 90.13 | 4.23x10$^{-11}$ | 10.03 |

**Figure captions:**

**Fig.1**: (a) HRXRD pattren of all Sr doped La$_{2-x}$Sr$_x$CoMnO$_6$ (0 $\leq$ x $\leq$ 1) series of samples, (b) magnified view of the XRD peak at 2θ =32.5° for all samples, and (c) & (d) shows the Rietveld refinement of XRD pattern of x = 0 and x = 0.5 doped samples.

**Fig.2**: (a) Normalized XANES spectra at Mn L$_{2, 3}$-edges of the La$_{2-x}$Sr$_x$CoMnO$_6$ series of samples together with those of MnO, Mn$_2$O$_3$ and MnO$_2$ for reference, (b) Normalized XANES spectra at Co L$_{2, 3}$-edges of the La$_{2-x}$Sr$_x$CoMnO$_6$ samples

**Fig.3**: Raman spectra of (a) La$_2$CoMnO$_6$ and (b) La$_{2-x}$Sr$_x$CoMnO$_6$ (0 $\leq$ x $\leq$ 1) samples, and(c) & (d) shows the Sr content variation of Raman spectra and FWHM for A$_{1g}$ and B$_{1g}$ modes respectively.

**Fig. 4**: (a) Temperature dependent resistivity under zero magnetic field and (b) isothermal field variation of MR (%) at 110 K for La$_{2-x}$Sr$_x$CoMnO$_6$ (0 < x < 1) samples.

**Fig. 5: (a-f):** M (T) data under ZFC and FCC modes for 100 Oe field in La$_{2-x}$Sr$_x$CoMnO$_6$ (0 < x < 1) samples and the inset to their respective figures represents the first derivative of ZFC magnetization with temperature; the inset of Fig. 5(d) is the magnified view of the ZFC magnetization to illustrate the low-temperature state of PS.

**Fig. 6: (a-f):** Temperature dependent χ′ with different frequencies of 1 Oe ac signal for Sr doped La$_{2-x}$Sr$_x$CoMnO$_6$ (0 $\leq$ x $\leq$ 1) samplesand the inset to their respective figures shows power law fit to experimental data of T$_f$ vs.τ data for Sr doping of x = 0.1 to 0.5.

**Fig.7:** Phase diagram of La$_{2-x}$Sr$_x$CoMnO$_6$ (0 $\leq$ x $\leq$ 1) samples with Sr doping.

**Fig.8:** (a-e) Isothermal ZFC-M (H) measurements at 5 K for La$_{2-x}$Sr$_x$CoMnO$_6$ (x = 0, 0.25, 0.5, 0.75 and 1.0) samples, and (f): Sr variation of M$_S$, M$_r$ and H$_C$ at 5 K (left panel) and % of ASD (right panel).

**Fig.9:** The Sr content variation of ZEB and CEB effects at 5 K.

**Fig. 1**

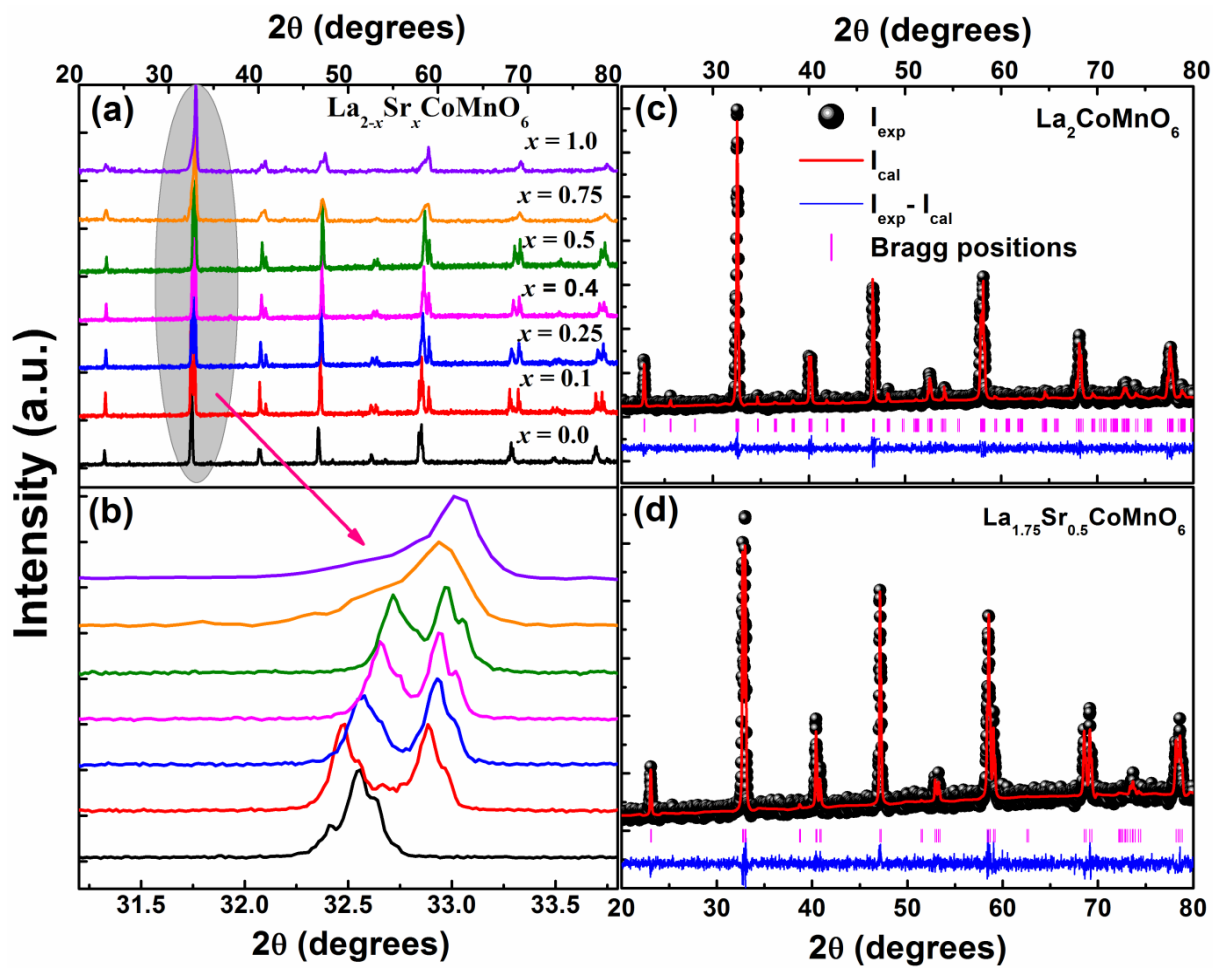

**Fig. 2**

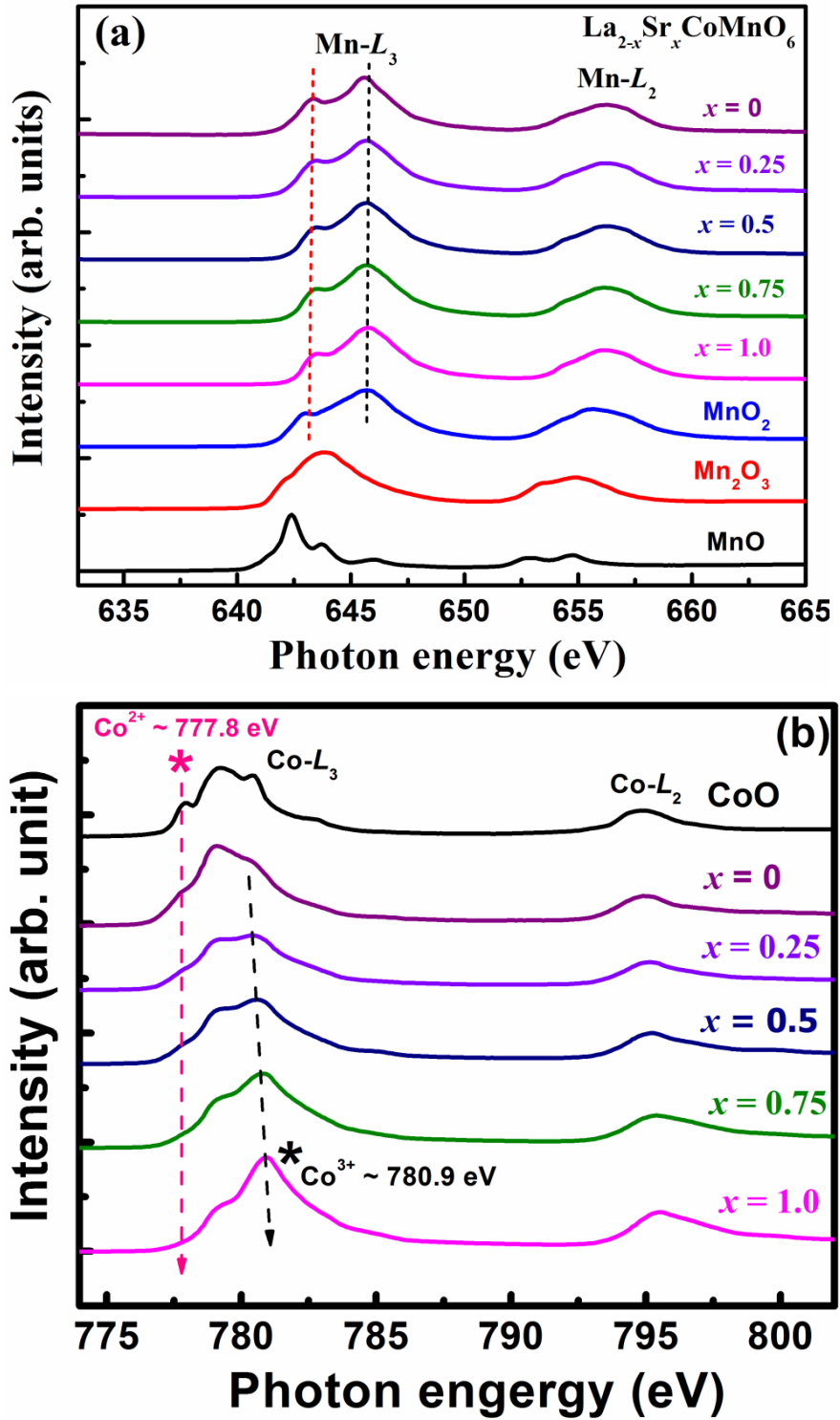

**Fig. 3**

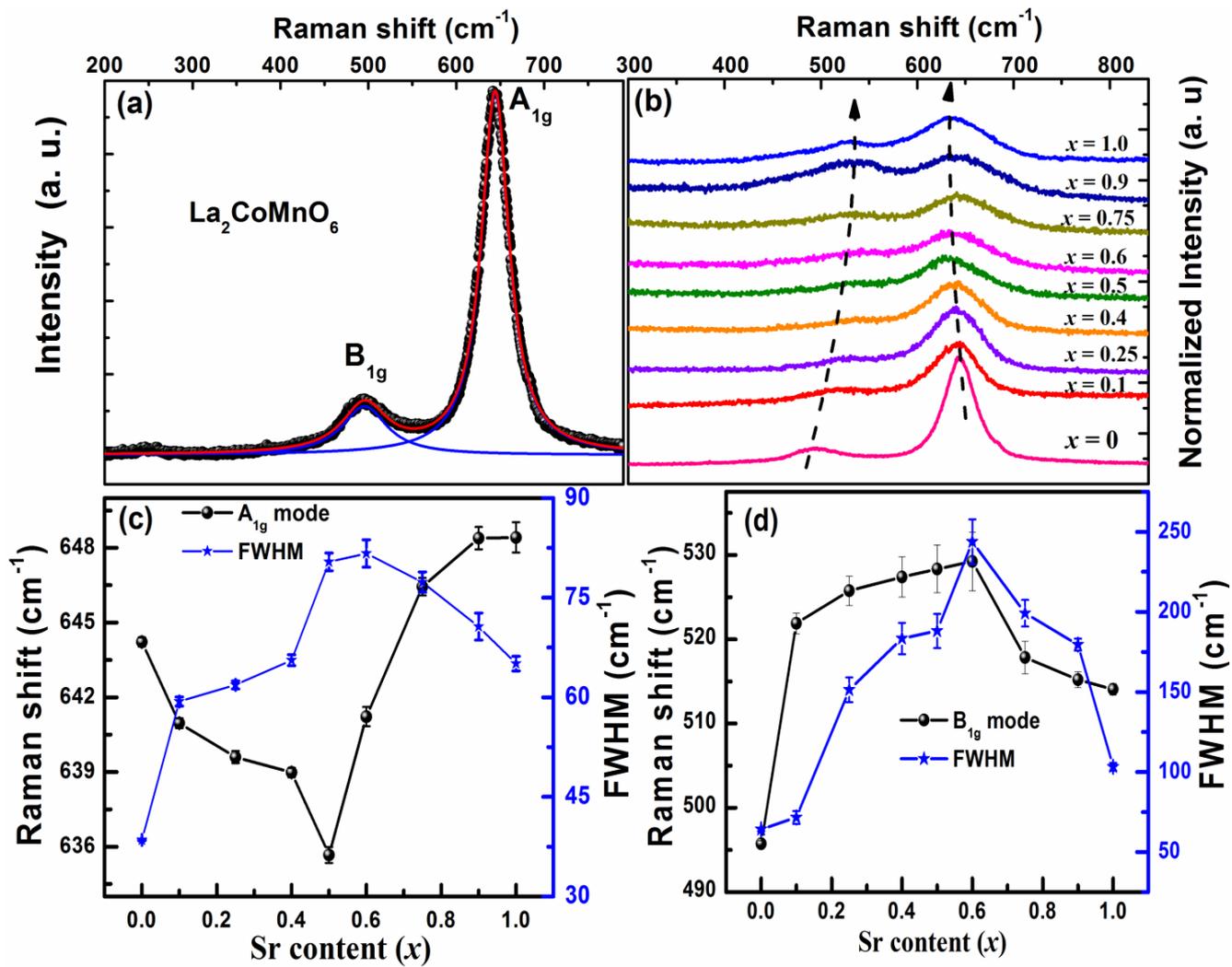

**Fig. 4**

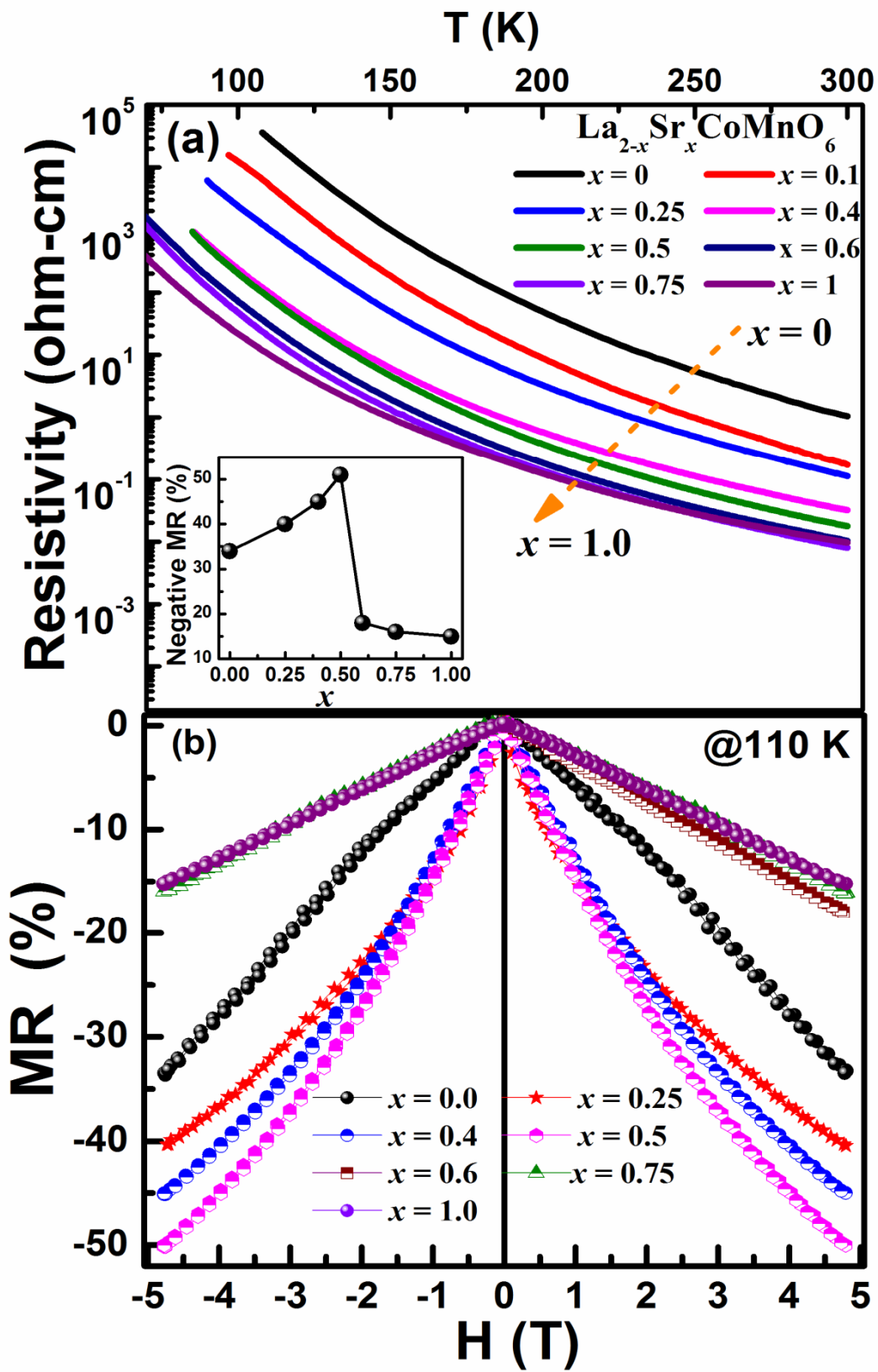

**Fig. 5**

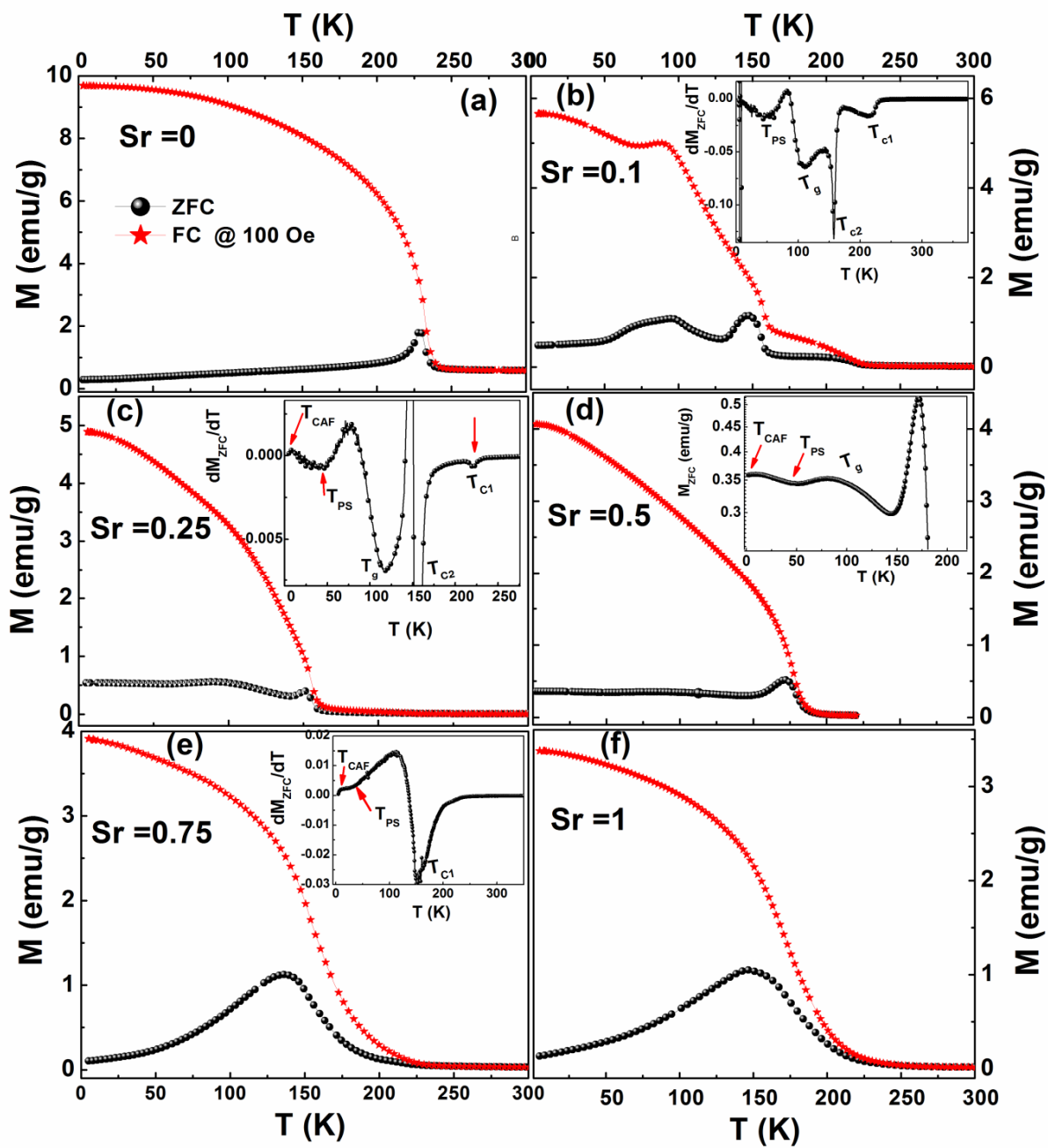

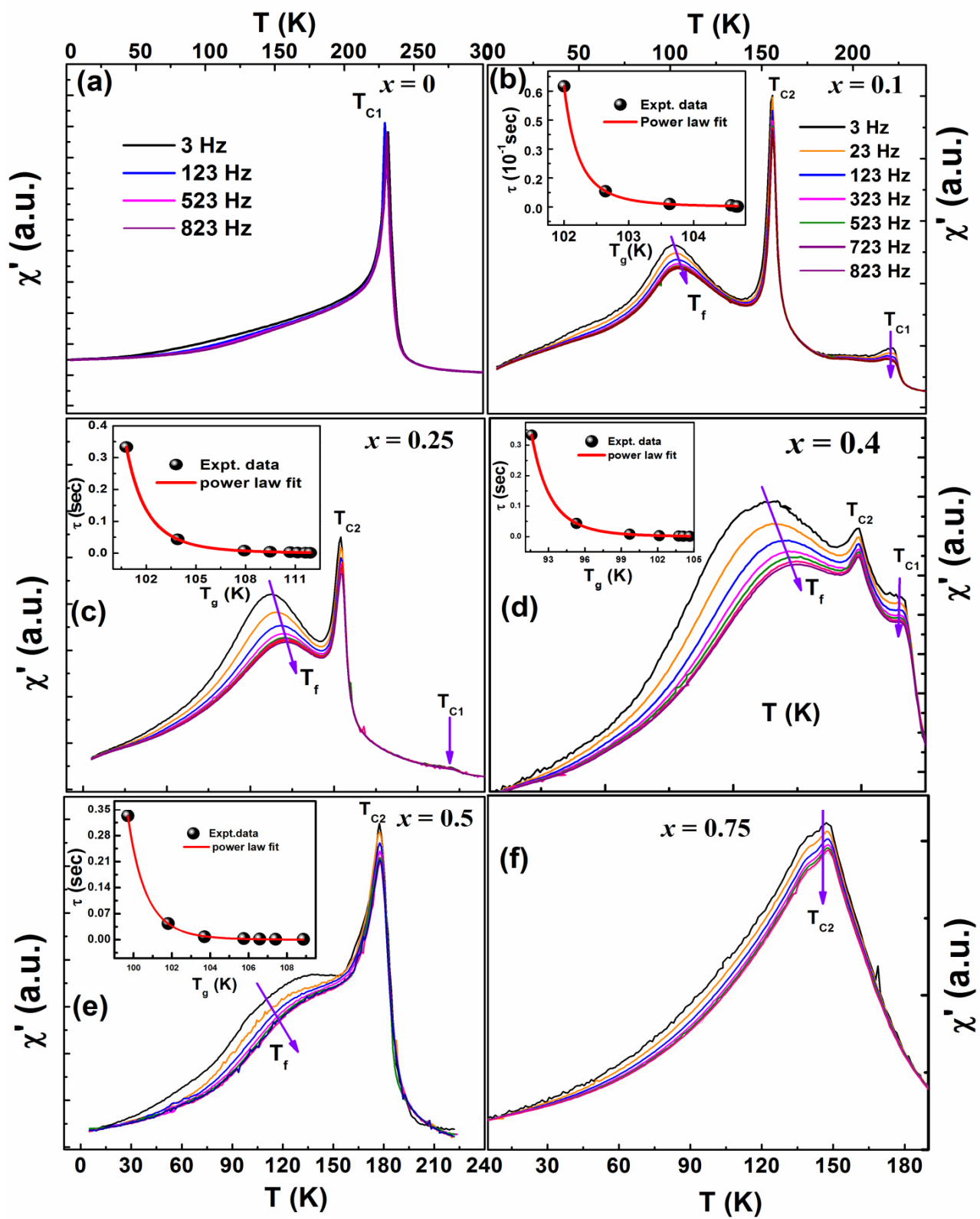

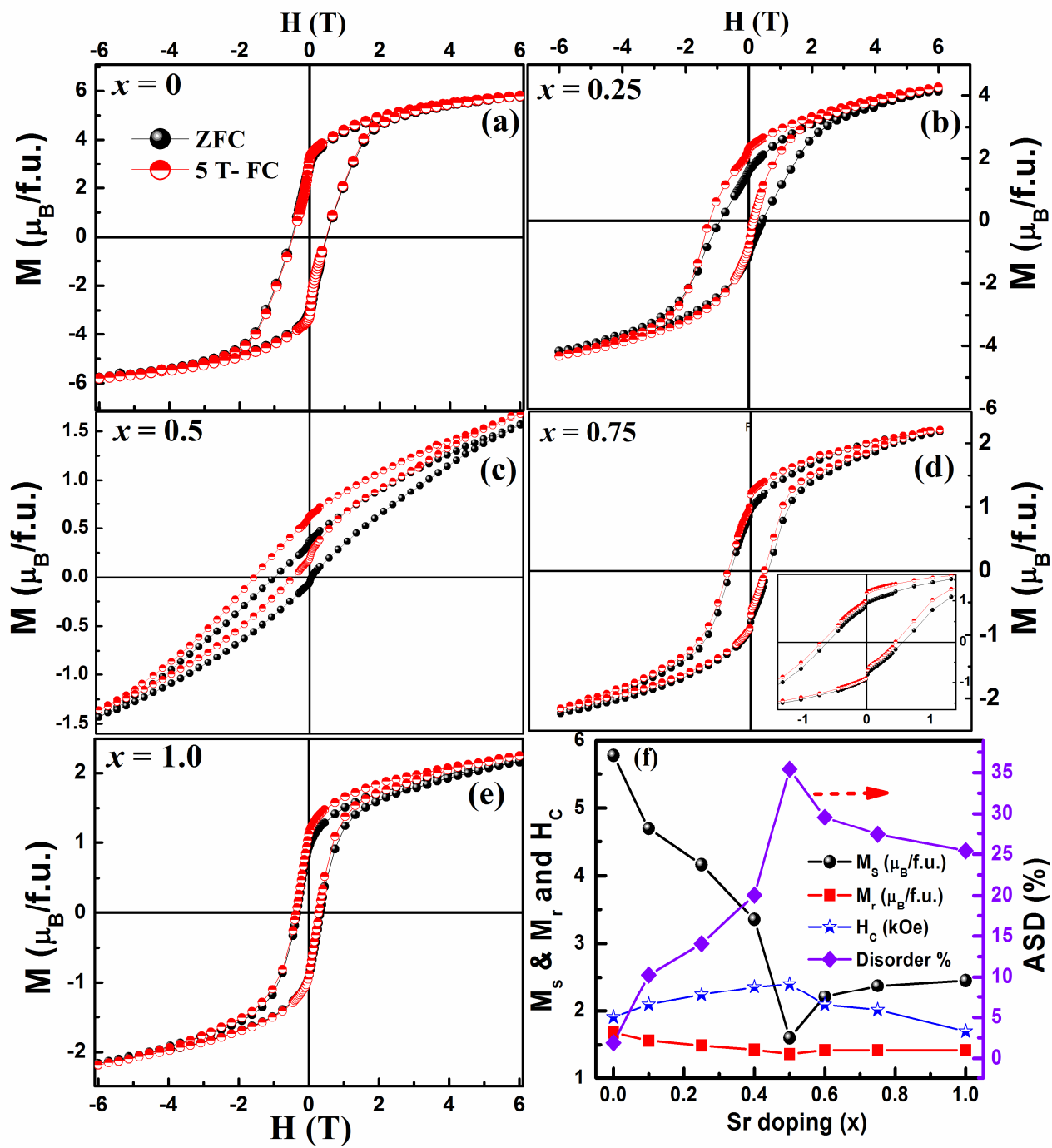

**Fig. 9**

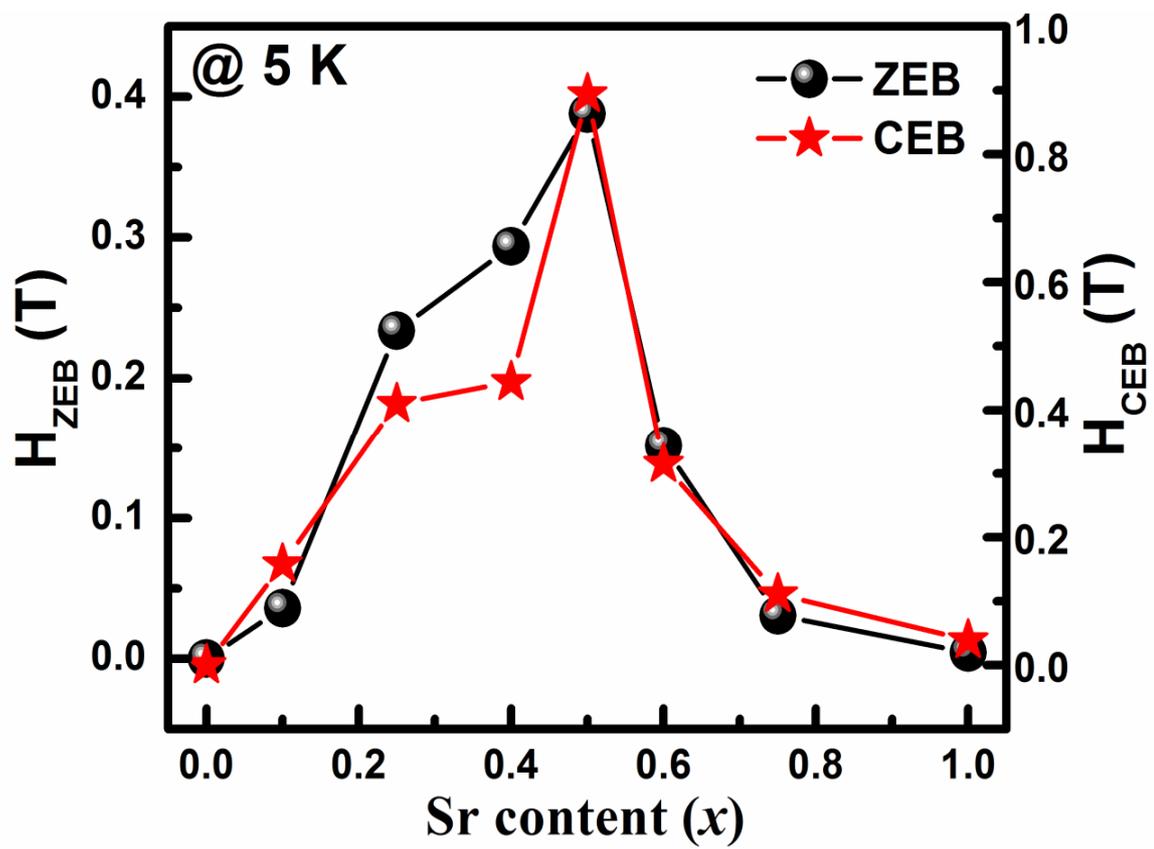